\documentclass[%
 aip,
 amsmath,amssymb,
 preprint,%
]{revtex4-2}

\usepackage{graphicx}
\usepackage{dcolumn}
\usepackage{bm}
\usepackage{color}
\usepackage{hyperref}
\usepackage{CJK}

\begin{document}
\begin{CJK*}{UTF8}{gbsn} 

\title[droplet effect on shear thinning]{Shear thinning of non-Brownian suspensions and its variation at different ambient conditions} 

\author{Yuan Lin}
\affiliation{Institute of Ocean Engineering and Technology, Ocean College, Zhejiang University, Zhoushan 316021, China}
\affiliation{The Engineering Research Center of Oceanic Sensing Technology and Equipment, Ministry of Education, Zhoushan 316021, China}%
\affiliation{Hainan Institute, Zhejiang University, Sanya 572025, China}%

\author{Peiwen Lin}
\affiliation{Institute of Ocean Engineering and Technology, Ocean College, Zhejiang University, Zhoushan 316021, China}%

\author{Ying Wang}%
\affiliation{Institute of Ocean Engineering and Technology, Ocean College, Zhejiang University, Zhoushan 316021, China}%

\author{Jiawang Chen}
\email{arwang@zju.edu.cn}
\affiliation{Institute of Ocean Engineering and Technology, Ocean College, Zhejiang University, Zhoushan 316021, China}
\affiliation{The Engineering Research Center of Oceanic Sensing Technology and Equipment, Ministry of Education, Zhoushan 316021, China}%
\affiliation{Hainan Institute, Zhejiang University, Sanya 572025, China}%

\author{Zhiguo He}
\email{hezhiguo@zju.edu.cn}
\affiliation{Institute of Port, Coastal and Offshore Engineering, Ocean College, Zhejiang University, Zhoushan 316021, China}

\author{Thomas P{\"a}htz}
\affiliation{Institute of Port, Coastal and Offshore Engineering, Ocean College, Zhejiang University, Zhoushan 316021, China}

\author{Nhan Phan-Thien (Phan Thi{\d{\^e}}n Nh{\^a}n)}
\affiliation{Department of Mechanical Engineering, National University of Singapore, Singapore, 117575}%

\date{\today}

\begin{abstract}
Immiscible contaminants are commonly involved in naturally occurring suspensions. The resulting variations of their flow behavior has rarely been evaluated. Here, we investigate the variation of the viscosity of the oil-based two-phase suspension over a period of two years, which is exposed to the ambient air at the production stage. We find that the air's absolute humidity, which strongly varies with the seasons, causes exchanges of water droplets with the suspension, substantially altering its shear-thinning behavior. Only in winter, when the humidity is low, is the latter close to that of ideal two-phase suspensions. Our measurements suggest that, when the surface roughness of the suspended solid particles is sufficiently low, immersed droplets remain in a free state, effectively increasing repulsion between particles, weakening shear thinning. In contrast, when the roughness is sufficiently high, immersed droplets become trapped on the particle surfaces, inducing an attractive particle interaction via water bridging, enhancing shear thinning.%
\end{abstract}

 
\maketitle
\end{CJK*}

\section{INTRODUCTION}
Flow of dense suspensions with solid particles is ubiquitous in nature and industry. Such systems include crude oil, mud, drilling fluid, blood, concrete, liquid food, nanocomposite, to name a few. Even the ideal suspensions, which consist of rigid and spherical particles and Newtonian fluid, display complex flow behavior. This is controlled by the Peclet number $\text{Pe} = \dot{\gamma}a^2/D$, where $D = kT/(6\pi \eta_0 a)$ is the self-diffusivity of a particle of radius $a$ suspended in a fluid of viscosity $\eta_0$, and $kT$ is the Boltzmann temperature. With increasing shear rate $\dot{\gamma}$, or $\text{Pe}$, their viscosity first decreases (shear thinning) and then increases (shear thickening).~\cite{Brady1997,Foss2000,Stickel2005,Morris2009,mewis_wagner_2011,Lee2021,Safa2019,Blair2022} The shear thinning is usually attributed to the deformation of equilibrium structures stabilized by the Brownian motion (i.e., the random oscillation of suspending particles due to their collisions with liquid molecules) and therefore occurs at a low $\text{Pe}$. However, also for non-Brownian particle suspensions ($\text{Pe}\gg1$), significant shear thinning is frequently observed at low shear rates,~\cite{Singh2003,Dai2013,Lin2014,Denn2014,Tanner2018} the mechanism behind which is still an open issue. Particle interactions, such as particle adhesion and friction, is a likely candidate.~\cite{Richards2020,Boyer2011,Trulsson2012} For example, it has been proposed that decreasing friction between particles with rough surfaces with increasing shear stress gives rise to shear thinning.~\cite{Guazzelli2018,Chatte2018,Gomez2020,Lobry2019,Arshad2021,Blanc2018,Comtet2017} Alternatively, it has been pointed out that suspended particles may form clusters in shear flow, and their break-up with increasing shear rate may cause shear thinning behavior.~\cite{Papadopoulou2020,Zhang2021,Gilbert2022} In fact, the organization of particles is a strain-dependent process: in shear flow with a low shear rate, particles approach each other and consequently clusters are formed at the equilibrium state when a critical strain is achieved.~\cite{Lin2015,Lin2021,Lin2021a} 

The degree to which shear thinning occurs substantially varies between experimental studies,~\cite{Lin2014_1,Zarraga2000,Dai2014,Haleem2009,Papadopoulou2020} with the origin being still an open issue. While the larger friction between rougher particles generally tends to enhance particle interactions and therefore shear thinning, it may also reduced them via preventing close particle contacts,~\cite{Gallier2014} thus hindering the formation of clusters.~\cite{Lin2021a} Furthermore, immiscible contaminants, such as air bubbles and water droplets much smaller than the suspended particles, can become trapped in laboratory-made suspensions because only rarely, if at all, have such suspensions been heard to be degassed or produced under dehumidified ambient-air conditions. Such contaminants, once formed, can be stable in the system for quite a long time compared to particle-scale contaminants.~\cite{Ohgaki2010,Ball2012,Alheshibri2016,Gao2021,Sun2016,Fang2018,Ankur2016,McClements2012} and potentially drastically alter the flow behavior.~\cite{Georgiev2018,Truby2014,Birnbaum2021}

Here, we show that the flow behavior of such laboratory-made oil-based suspensions is strongly affected by the absolute humidity of the ambient air. This is due to the involving of fine water droplets at the high-humidity condition, which alters the interaction between solid particles. This casts doubt on the conclusions of many studies employing such suspensions to investigate flow rheology.

\section{EXPERIMENT}

\begin{figure*}[tbh]
 \includegraphics[width=0.5\textwidth]{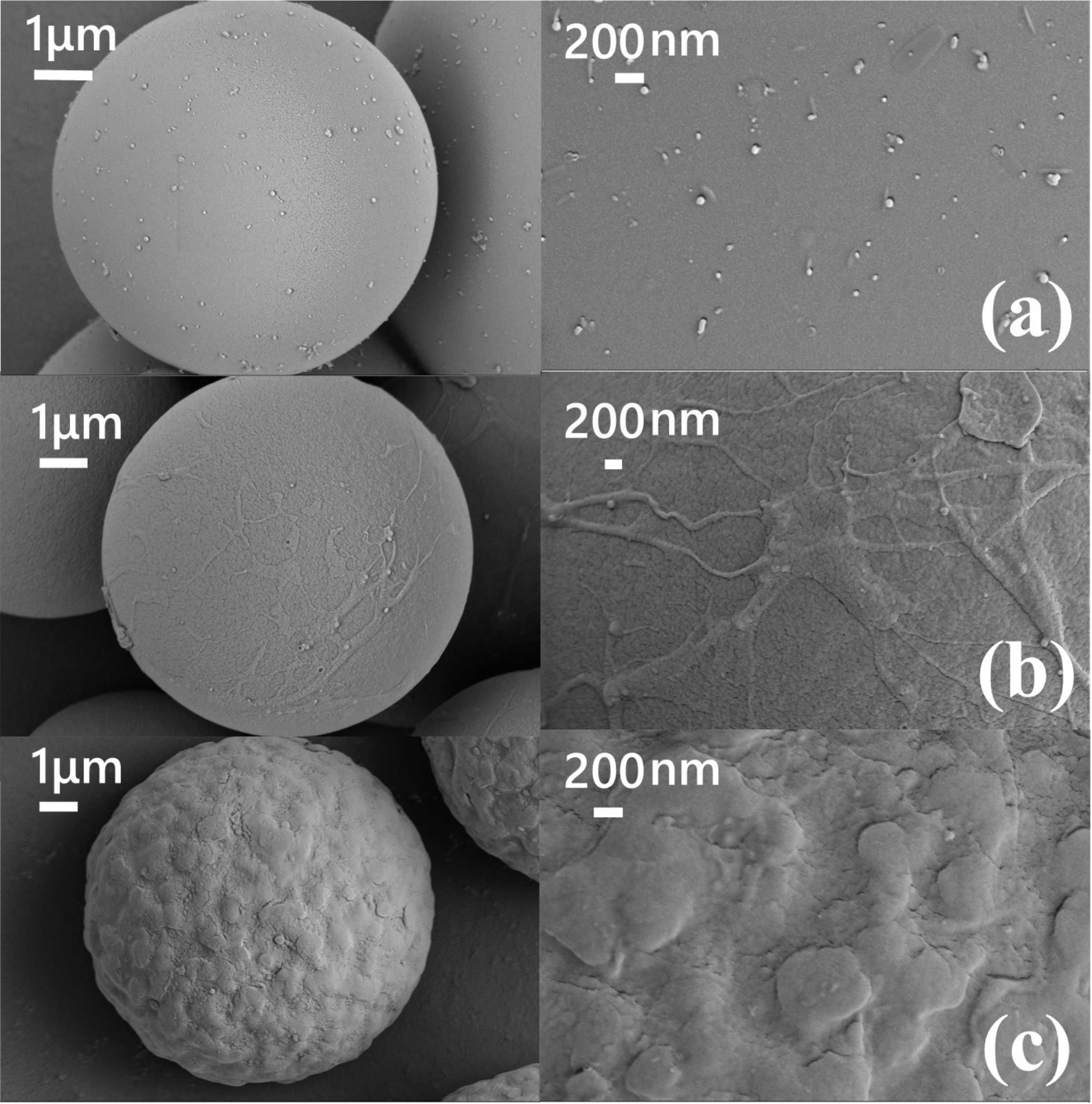}\\
  \caption{SEM images of (a) smooth glass particles, (b) moderate rough and (c) rough PMMA particles, adopted in the experiment.}\label{Fig1}
\end{figure*}

A silicone oil (polydimethylsiloxane) with a nearly constant viscosity, $\eta_0 = 0.51\;\text{Pa}\cdot\text{s}$ at $\dot{\gamma} \leq 100\;\text{s}^{-1}$, was adopt as the liquid phase of the suspensions used in this study. The density of the fluid is $0.97\;\text{g}/\text{cm}^3$ at $25\;^\circ \text{C}$. For the solid phase, we use three kinds of particles, namely two types of PMMA (polymethyl methacrylate) particles with different surface roughness, as well as smooth and hollow particles made of the borosilicate glass. All these particles have the average diameter of $10\;\mu \text{m}$. Glass particles are polydispered in size, and are basically smooth with small debris absorbed on the particle surface, as shown by the SEM (Scanning Electron Microscope) images in Fig.\ref{Fig1} (a). The two types of PMMA particles have the roughness estimated to be $O(10)\;\text{nm}$ and $O(10^2)\;\text{nm}$, respectively, as shown in Fig.\ref{Fig1} (b) and (c). The density of the hollow glass particles and the PMMA particles are $1.1$ and $1.17\;\text{g}/\text{cm}^{3}$, respectively. The surface energy of the two solid materials are similar (about $40\;\rm mN/m$).~\cite{Szymczyk2008,HU2015,Szymczyk2011} 

\begin{figure*}[tbh]
 \includegraphics[width=0.7\textwidth]{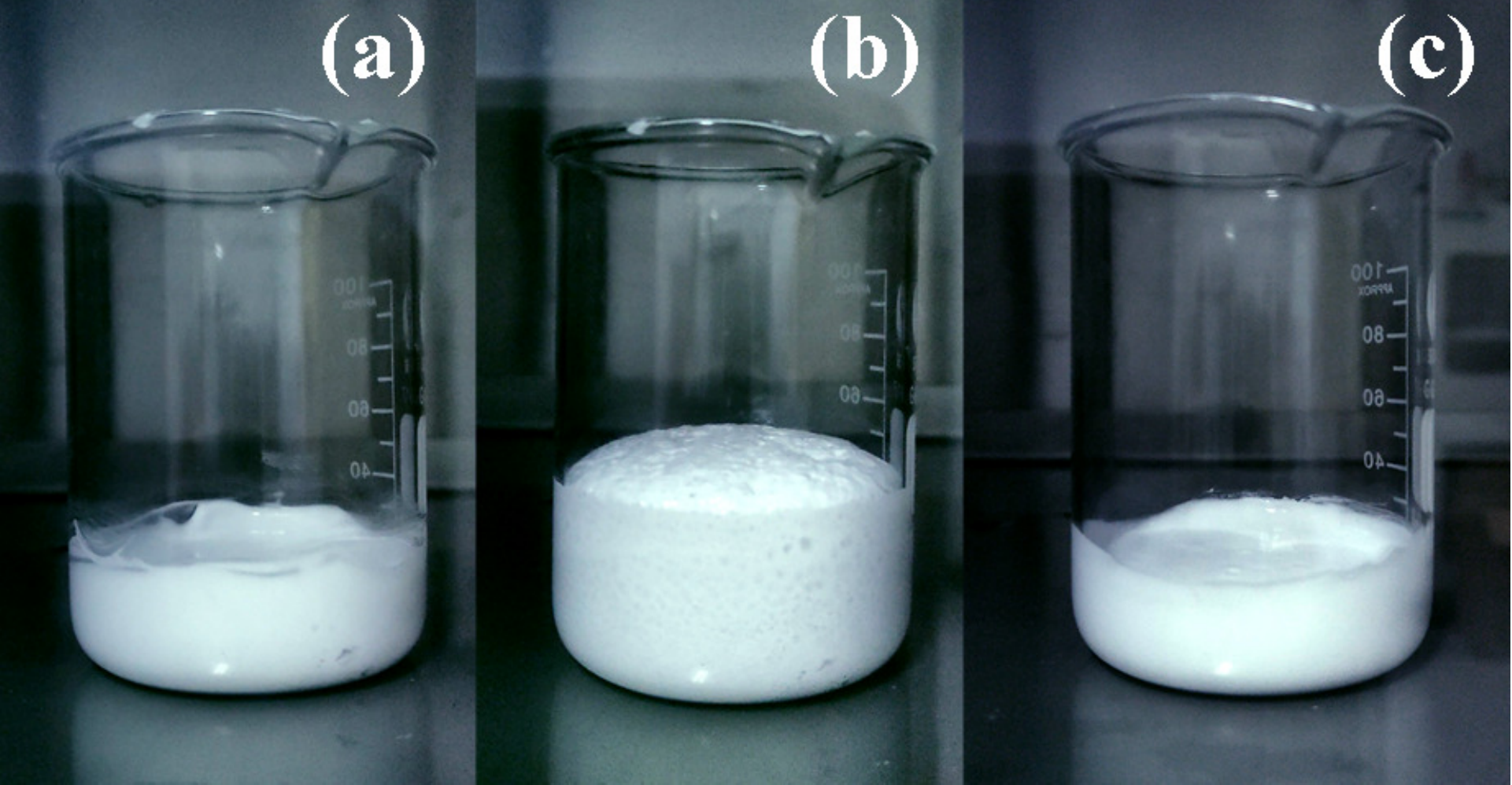}\\
  \caption{Degassing of the suspension with hollow glass particles at the stage (a) before degassing, (b) during degassing and (c) after 10 hrs' rest in the vacuum oven. The volume fraction of the suspension is $40\;\%$.}\label{Fig2}
\end{figure*}

During the sample preparation, the room temperature is kept at around $25\;^\circ\text{C}$. The particle powder is homogenized with the silicone oil, after which different treatments are applied to investigate the influence from air bubbles and the moisture involved. Mixtures are rested under the natural ambient condition at $25\;^\circ \text{C}$ for $24\;\text{hrs}$, during which large bubbles can be separated from the suspension by mismatch of densities, while fine (probably nano-scaled) bubbles and droplets remain in the suspension. Consequently, samples containing the above mentioned third-phase components are obtained. On the other hand, to approach the strict two-phase suspensions as the control group, mixtures are degassed in a vacuum oven for $24\;\text{hrs}$ at $25\;^\circ \text{C}$ to remove both the air and the moisture in the system. During this degassing process, we observe a large amount of bubbles grow and finally escape through the free surface at the early stage, as shown in Fig.\ref{Fig2}. All visible bubbles disappear within $12\;\text{hrs}$. 

In the following, G1 and G2 are referred to the non-degassed and degassed suspensions with glass particles, and P1 and P2 are referred to the non-degassed and degassed suspensions with PMMA particles, respectively. 

In the rheological experiment, we adopt a DHR-1 rotational rheometer (TA Instrument, US) with a cone-plate geometry with the diameter of $20\;\text{mm}$. The upper plate has a cone angle of $4\;^\circ$, and is truncated $109\;\mu\text{m}$ at the vertex to avoid the jamming effect by particles. We adopt a $45\;\text{s}$ pre-shearing procedure with a shear-rate, $\dot{\gamma}=10\;\text{s}^{-1}$ to ensure the same shear history for all the samples. We perform a ramp-up shearing test using the shear-rate-control mode to measure the viscosity as a function of the shear rate. The shearing time is $5\; \text{s}$ in each shearing step. Furthermore, a start-up shearing test with a constant shear rate is performed to investigate the transition behavior of the viscosity before the equilibrium state. Afterwards, an oscillatory shear test using the stress-control amplitude sweep mode is carried out to investigate the structure formed in the previous start-up shear test. The experimental temperature is kept at $25\;^\circ\text{C}$. The boundary effect such as boundary slip is negligible based a gap-dependent test using a $25\;\text{mm}$ parallel-plate geometry, and therefore is not considered in our study.

\section{RESULTS AND DISCUSSION}

\begin{figure*}[tbh]
 \includegraphics[width=0.6\textwidth]{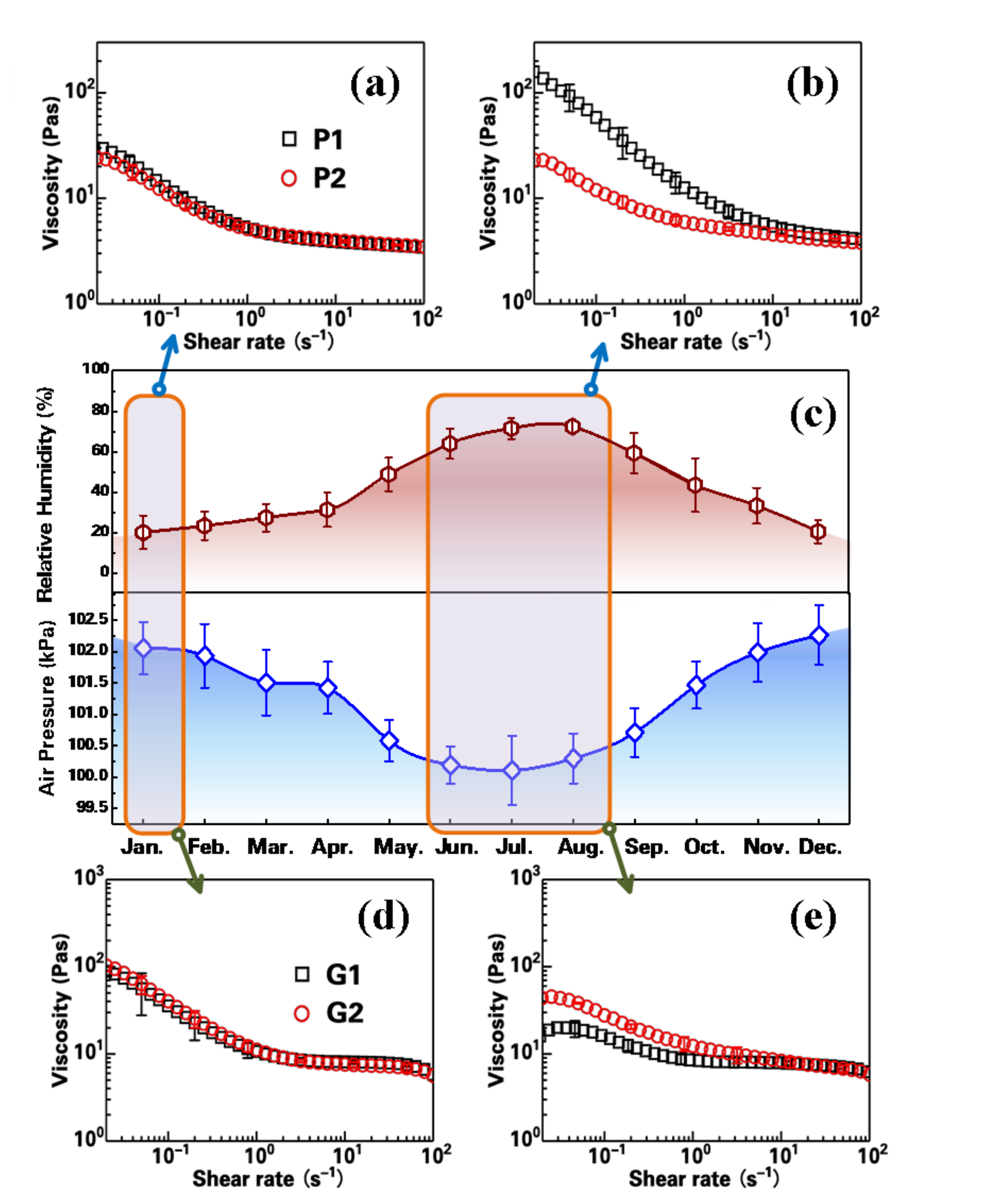}\\
  \caption{(a) and (b) viscosity versus shear rate curves captured in different periods for P1 and P2 samples, respectively; (c) variation of the monthly averaged air pressure and relative humidity (converted to the equivalent relative humidity at $25\;^\circ \text{C}$) from January to October; (d) and (e) viscosity versus shear rate curves captured in different periods for G1 and G2 samples, respectively. The volume fraction of samples are $40\;\%$. The rheologcial data is collected over the year 2020 and 2021 and therefore, the pressure and humidity data in (c) are also averaged over these two years. Each curve in (b) and (e) was averaged over four tests carried out in June, July and August, respectively.}\label{Fig3}
\end{figure*}

\begin{figure*}[tbh]
 \includegraphics[width=0.8\textwidth]{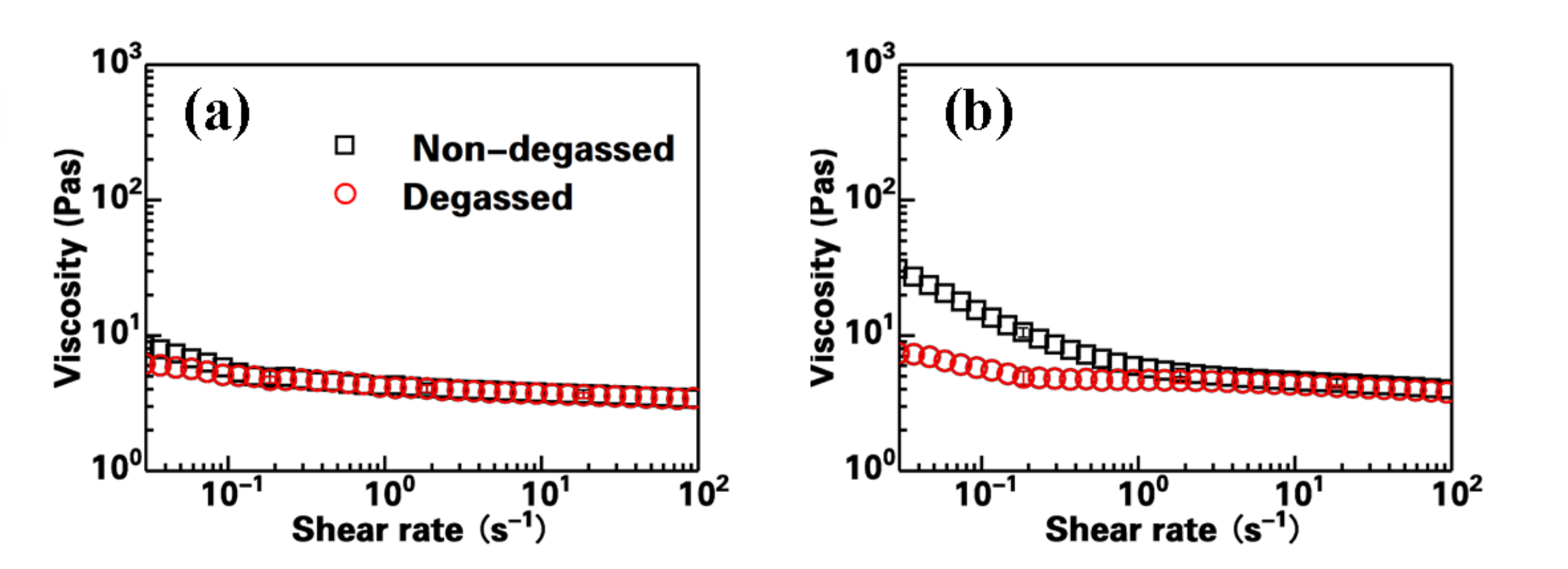}\\
  \caption{Viscosity versus shear rate curves for suspensions with very rough PMMA particles at the volume fraction of $40\;\%$ captured in (a) January and (b) June to August.}\label{Fig4}
\end{figure*}

Considerable shear thinning can be observed for all the samples in our study. As shown in Fig.\ref{Fig3}, it is observed that the shear thinning behavior for non-degassed samples changes considerably, depending on when the test is applied in a year. By contrast, for the degassed samples, the shear thinning behavior is nearly invariable. From Fig.\ref{Fig3} (b) and (e), it can be found that, during June to August, the shear thinning is less obvious for G1 compared to the degassed counterpart (G2), and at the high-shear-rate range, the two viscosity curves overlap, while the shear thinning behavior of P1 is more significant compared to P2. As shown in Fig.\ref{Fig4} (b), the phenomenon is similar for suspensions with rougher PMMA particles (with roughness estimated to be $O(10^2) \;\rm nm$, see Fig.\ref{Fig1} (c)). On the other hand, if the experiment is carried out in January, as shown in Fig.\ref{Fig3} (a) and (d), as well as Fig.\ref{Fig4} (a), all the shear thinning behavior for non-degassed suspensions approach the degassed counterpart. From August to January, both the barometric pressure and the humidity changes considerably according to the data from the local weather bureau, as shown in Fig.\ref{Fig3} (c). One or both of these factors should account for the altering of the shear thinning behavior for the non-degassed samples. For suspensions with smooth glass particles, as shown in Fig.\ref{Fig3} (d) and (e), a second shear thinning can be observed at the shear rate close to $100\;s^{-1}$, which is due to the shear thinning of the fluid matrix.~\cite{Lin2014_1,Lin2017} In shear flow, suspending particles approaches each other, giving rise to the lubricative interaction. This leads to a local flow enhancement between particles, which consequently shifts the onset of shear thinning of the fluid phase to a lower bulk shear rate. For suspensions with rough particles (PMMA particles in this study), as shown in Fig.\ref{Fig3} (a) and (b) and Fig.\ref{Fig4}, the second shear thinning disappears. This agrees with the finding in the previous study.~\cite{Lin2021a} It is considered that asperities on rough particles hinders the further approaching between particles, reducing the lubrication force between particles and thereby the local flow enhancement. Consequently, the second shear thinning is suppressed.

\begin{figure*}[tbh]
 \includegraphics[width=0.8\textwidth]{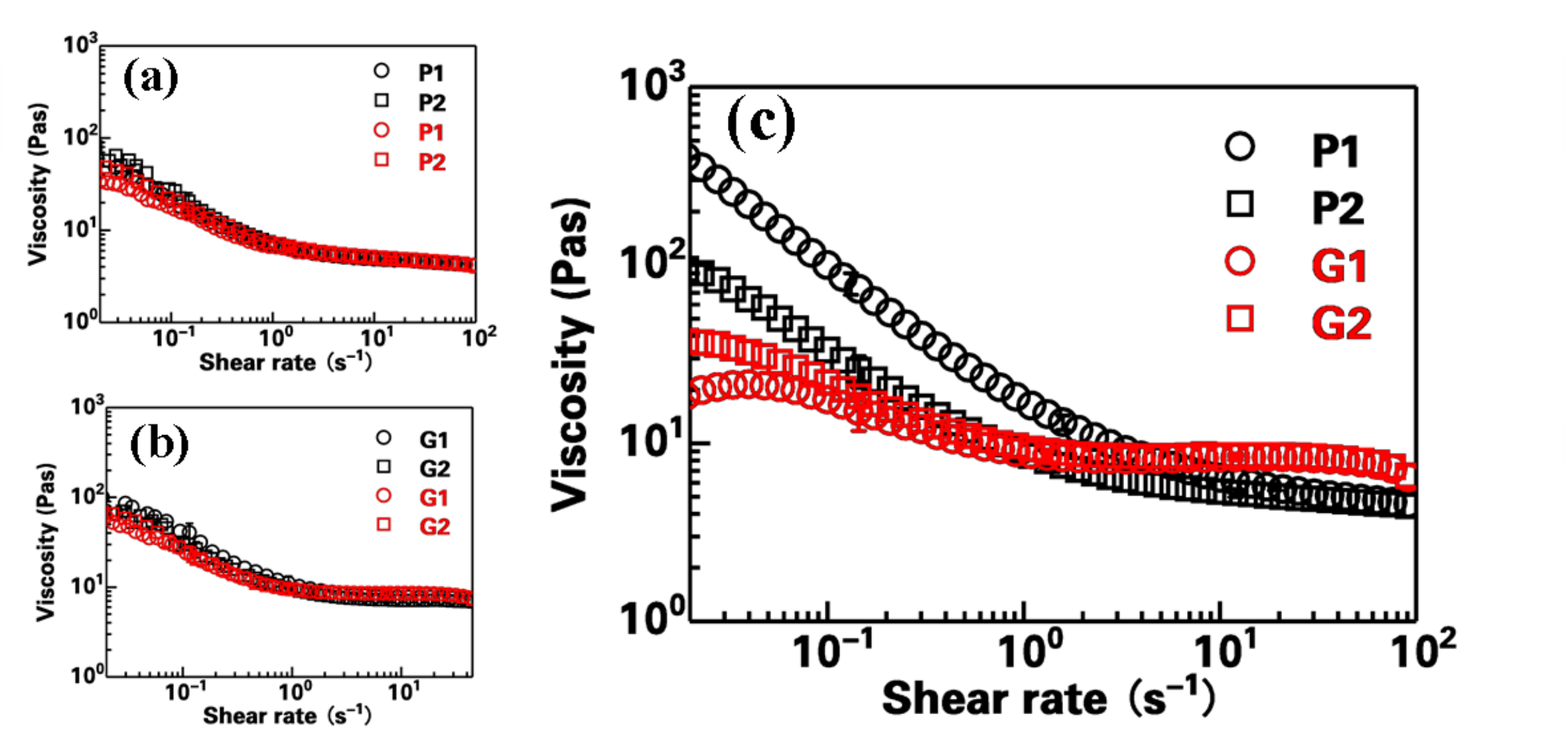}\\
  \caption{Viscosity curves measured in the city of Chengdu (black symbols) and Lanzhou (red symbols) with the air pressure of $96.21\;\rm kPa$ and $85.32\;\rm kPa$, respectively, for suspensions with (a) PMMA particles and (b) glass particles {at the volume fraction of $40\;\%$}. The relative humidity at $25\;^\circ\text{C}$ are below $30 \%$ in both cities. (c) Viscosity curves measured under an artificial relative humility above $75 \%$ at $25\;^\circ\text{C}$ with the air pressure above $102\;\rm kPa$.}\label{Fig5}
\end{figure*}

\begin{figure*}[tbh]
 \includegraphics[width=0.6\textwidth]{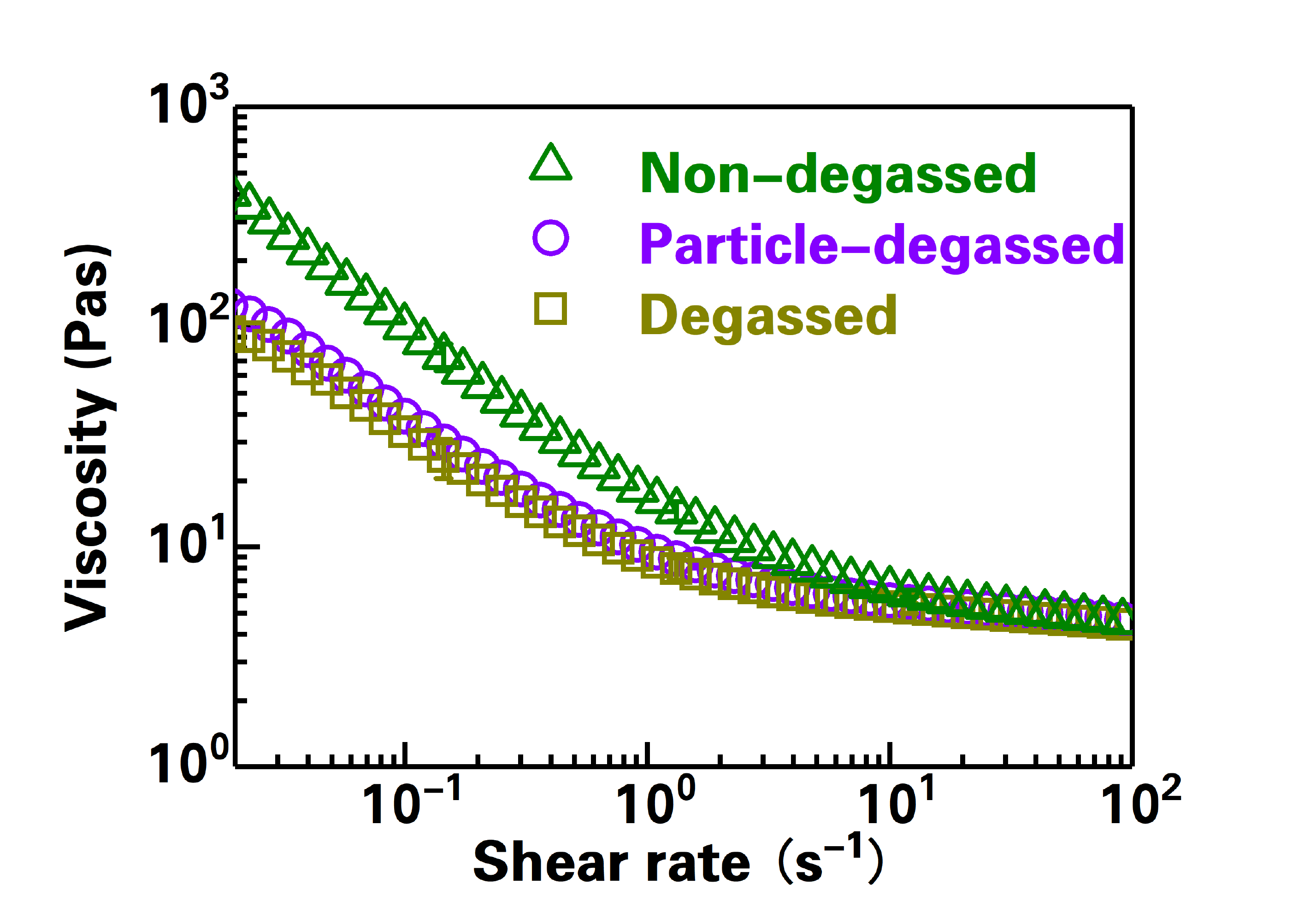}\\
  \caption{Viscosity versus shear rate curves for non-degassed and degassed suspensions with PMMA particles captured in January at an artificial relative humidity above $75\%$ at $25\;^\circ\text{C}$. The viscosity curve of suspension with dehumidified PMMA particles (circles) is also presented. The volume fraction of the suspension is $40\;\%$.}\label{Fig6}
\end{figure*}

The key factor that affects the shear thinning are identified from the above mentioned two factors, namely the air pressure and ambient humidity. The influence from solely the air pressure is investigated by applying the rheological test in the city of Chengdu and Lanzhou, located in west and northwest part of China, respectively, where the altitude reaches $1720 \;\rm m$, and consequently the air pressure is as low as $85.32\;\rm kPa$. The relative humidity is below $30\%$ at the room temperature of $25\;^\circ\text{C}$. As shown in Fig.\ref{Fig5} (a) and (b), altering of the shear thinning due to the degassing process is not obvious at such low ambient pressure. Therefore, change of the ambient pressure should not account for the variation of the shear thinning behavior shown in Fig.\ref{Fig3} (b) and (e). To investigate the effect of the humidity, we increase the relative humidity of the ambient air from the natural $45\%$ to around $75\%$ using a humidifier, and the air pressure is kept at $102 \;\rm kPa$ (in January). The powder of suspending particles are exposed to such moist condition for $24\;\rm hrs$ before it is used for the preparation of suspensions. As shown in Fig.\ref{Fig5} (c), we manage to reproduce the variation of the shear thinning behavior observed in Fig.\ref{Fig3} (b) and (e). Moreover, if we use the sample that is prepared using the above-mentioned moist powder which alone is dehumidified in the vacuum oven ($24 \;\rm hrs$, $25\;^\circ\text{C}$), it is observed that shear thinning behavior approaches that of a degassed suspension (see Fig.\ref{Fig6}). We thereby confirm that the ambient humidity, rather than the air pressure, is the key factor affecting the rheology of the oil-based non-colloidal suspensions. At a high level of humidity, water droplets are involved in the suspension by moist particles. Such droplets, although with negligible volume fraction, alters the shear thinning behavior of the suspension. This is consistent with the finding by Koos et al., ~\cite{Koos2011,Koos2012} who discovered that by adding a small amount of water, the oil-based suspension can be turned from weakly elastic to gel-like.

\begin{figure*}[tbh]
 \includegraphics[width=0.7\textwidth]{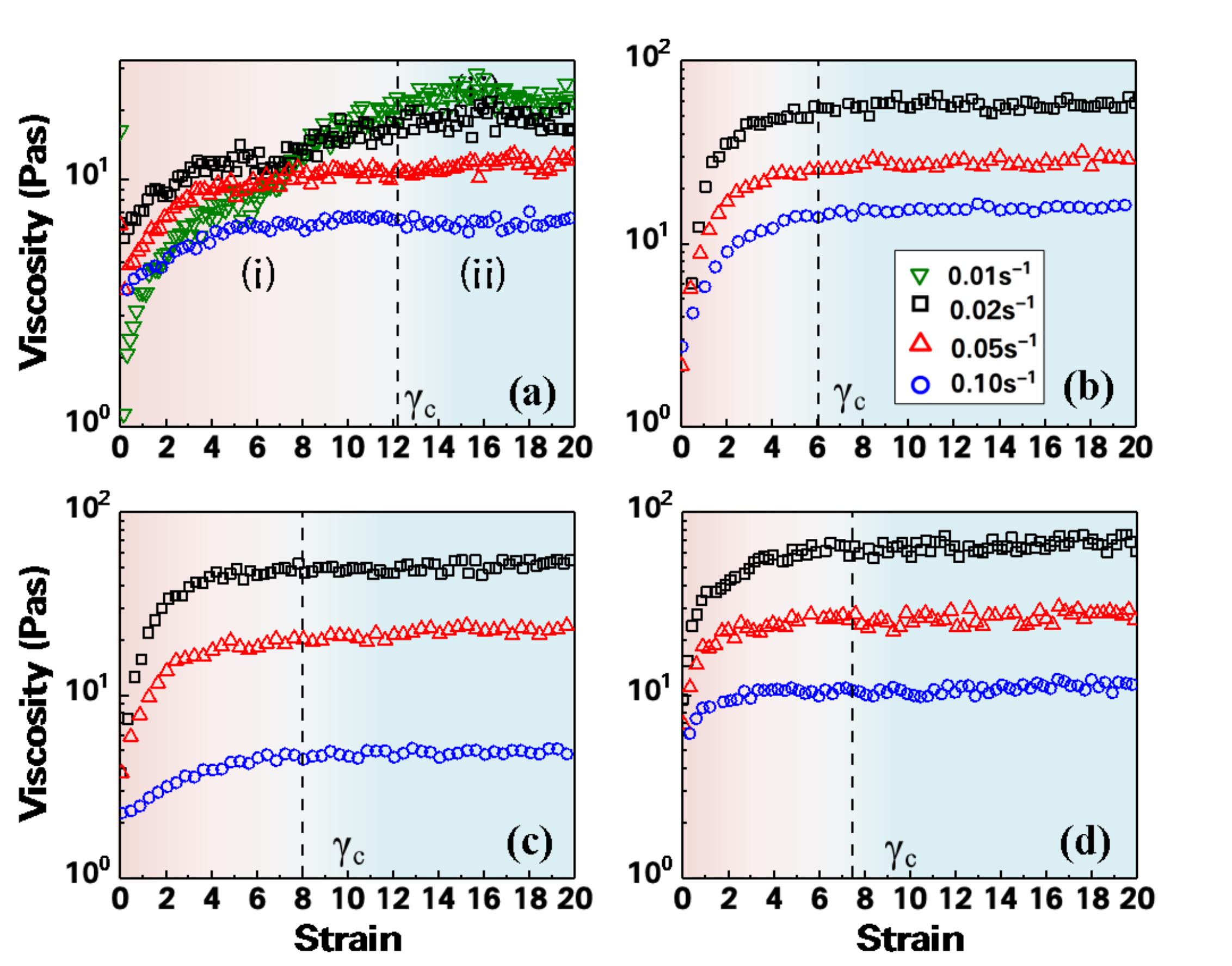}\\
  \caption{Evolution of the transitional viscosity with shear strain in the start-up shear flow at different shear rates for (a) G1, (b) G2, (c) P1 and (d) P2 samples, respectively, from (i) the transition state to (ii) the equilibrium state at $\phi=30\;\%$. Data are captured in the period during June to August.}\label{Fig7}
\end{figure*}

\begin{figure*}[tbh]
 \includegraphics[width=0.7\textwidth]{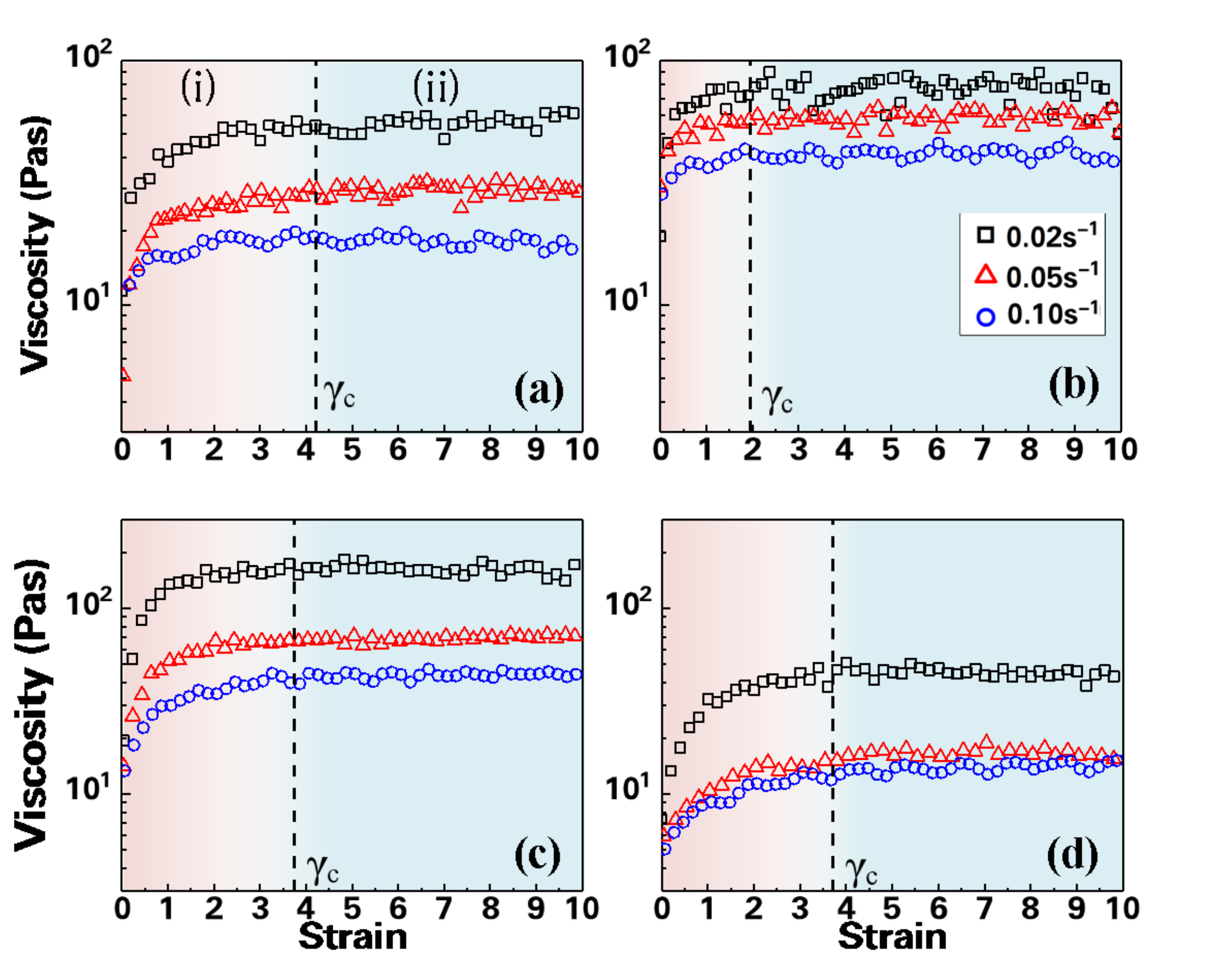}\\
  \caption{Evolution of the transitional viscosity with shear strain in the start-up shear flow at different shear rates for (a) G1, (b) G2, (c) P1 and (d) P2 samples, respectively, at $\phi = 40\;\%$. Data are measured in the period during June to August.}\label{Fig8}
\end{figure*}

\begin{figure*}[tbh]
 \includegraphics[width=0.7\textwidth]{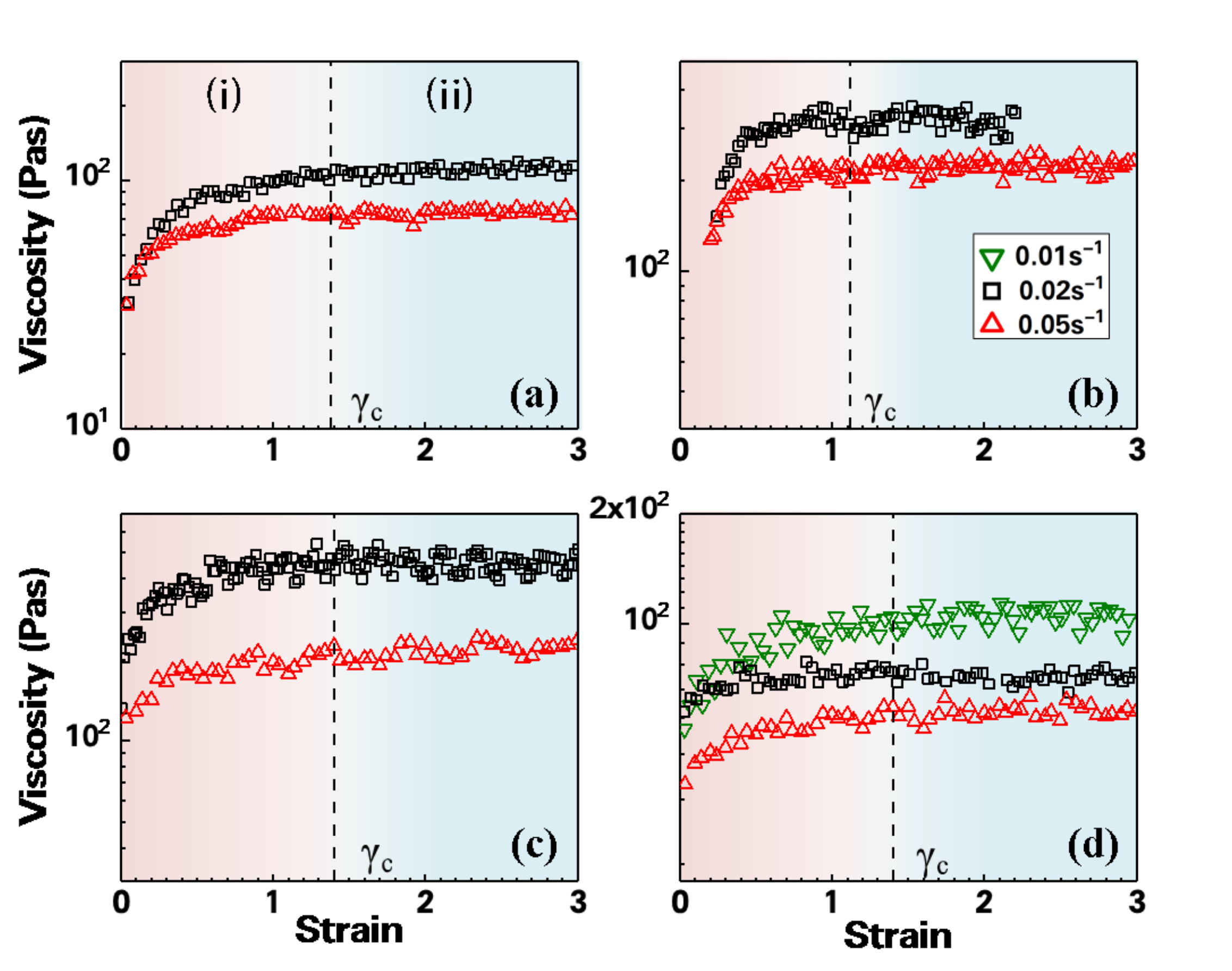}\\
  \caption{Evolution of the transitional viscosity with shear strain in the start-up shear flow at different shear rates for (a) G1, (b) G2, (c) P1 and (d) P2 samples, respectively, at $\phi = 50\;\%$. Data are measured in the period during June to August.}\label{Fig9}
\end{figure*}

\begin{figure*}[tbh]
 \includegraphics[width=0.7\textwidth]{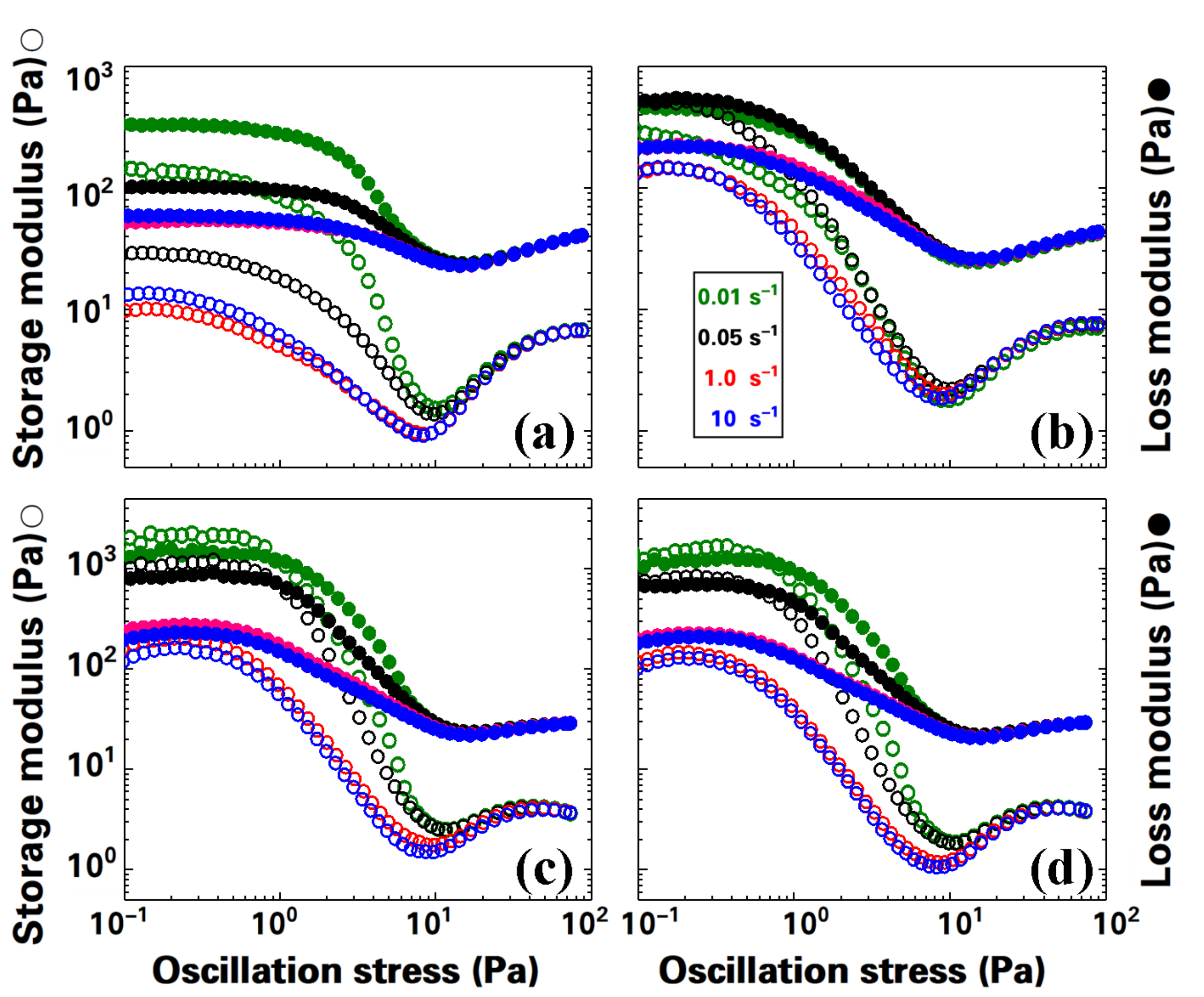}\\
  \caption{Evolution of the storage modulus, $G'$ and the loss modulus, $G''$ with the stress amplitude in the amplitude sweep test after the start-up shear test with various shear rates for (a) G1, (b) G2, (c) P1 and (d) P2 samples, respectively, at $\phi = 40\;\%$. The oscillation frequency is $1\;\text{Hz}$.}\label{Fig10}
\end{figure*}

It is proposed that shear thinning origins from the shear-induced organization of particles, during which shear-rate dependent clusters are formed.~\cite{Lin2015,Lin2021a} The particle aggregation is due to an adhesive particle interaction.~\cite{Richards2020,Ge2021} Therefore, it is essential that particles approach each other for the near-field adhesive interaction taking into account. Consequently, in a start-up shear flow, the approaching of particles from the initial state gives rise to the transition of the viscosity with increasing accumulated strain $\gamma$, as shown in Fig.\ref{Fig7}, Fig.\ref{Fig8} and Fig.\ref{Fig9}. The critical strain, $\gamma_c$, accounting for the equilibrium viscosity is observed to be independent of $\dot{\gamma}$. This agrees with the study by Lin et al.~\cite{Lin2015}. The forming of clusters can be confirmed by an oscillatory shear test following the start-up shearing test, in which the mode of stress-control amplitude sweep at the frequency of $1\;\text{Hz}$ is adopt. As shown in Fig.\ref{Fig10}, it is found that at the low-amplitude range, a higher plateau of dynamic modulus can be observed if a lower $\dot{\gamma}$ is applied in the previous start-up shearing test. Furthermore, the storage modulus, $G'$ becomes comparable to the loss modulus, $G''$, if $\dot{\gamma}$ is below $0.1\;\text{s}^{-1}$, which indicates a more solid-like property of the system. This reflects that a structure is formed in the previous shear flow with a low-enough $\dot{\gamma}$. Furthermore, the plateau $G'$ and $G''$ decrease with increasing $\dot{\gamma}$, indicating that the scale of the structure decreases with increasing $\dot{\gamma}$. This confirms that the shear thinning observed is due to the forming of shear-rate dependent clusters. 

\begin{figure*}[tbh]
 \includegraphics[width=0.8\textwidth]{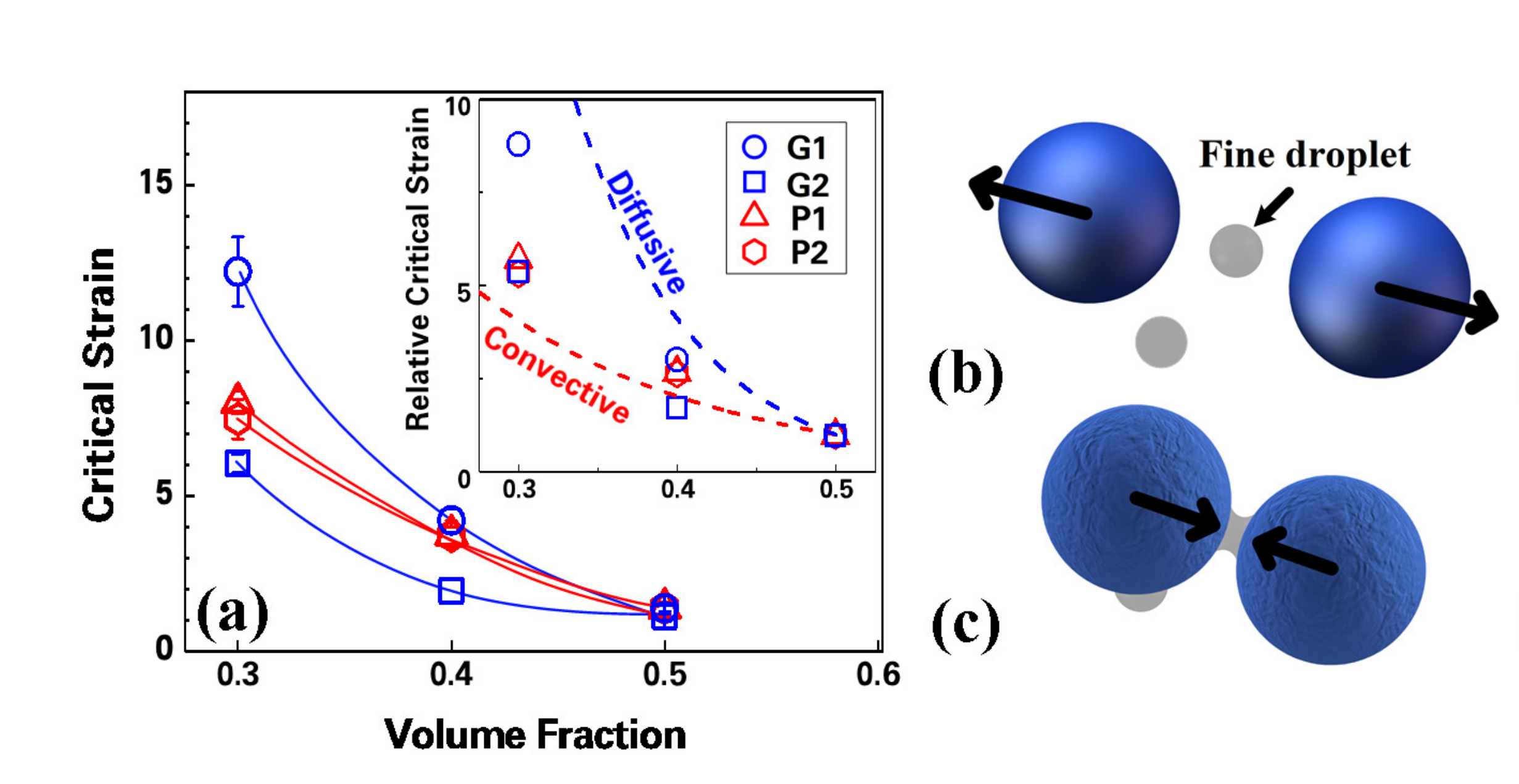}\\
  \caption{(a) Evolution of the critical strain, $\gamma_c$, with solid volume fraction. The inset shows the evolution of the relative $\gamma_c$ normalized by their respective value at $\phi = 50\;\%$. The red and blue dashed line indicates the evolution of $\gamma_c$ following the convective motion ($\gamma_c \sim \left\langle \varepsilon \right\rangle$) and diffusive motion ($\gamma_c \sim \left\langle \varepsilon \right\rangle^2$), respectively. (b) and (c) Illustration of the interaction between suspending particles and the fine droplet for smooth and rough suspending particles, respectively. Arrows represent the property (repulsive or attractive) of the equivalent inter-particle force between particles due to the presence of third-phase droplets.}\label{Fig11}
\end{figure*}

Based on the above mentioned mechanism of the shear thinning behavior, we propose that during the organization process, particles need to travel a relative distance of the average gap between two nearest suspending particles, $\left\langle \varepsilon \right\rangle$. The motion of particles before contact with their neighbors (i.e., at the transition state shown in Fig.\ref{Fig7} - Fig.\ref{Fig9}) is either ``convective'' or ``diffusive''. In the convective limit, particles moves along with streamlines, and $\gamma_c$ may be roughly estimated as $\gamma_c \sim \left\langle \varepsilon \right\rangle / \left\langle \delta \right\rangle$, where $\left\langle \delta \right\rangle$ is the vertical component (along the velocity-gradient direction) of the average distance between centers of nearest particles, and is considered to depend on the shearing history, which thus is insensitive to $\phi$.~\cite{Lin2021} Therefore, $\gamma_c \sim \left\langle \varepsilon \right\rangle$. In the diffusive limit, considering the diffusivity of particles is approached by the self-diffusivity $D_s = d(\phi)\dot{\gamma}\left\langle a \right\rangle ^2$, where $d(\phi)$ is assumed to be insensitive to the particle volume fraction, $\phi$, at a high particle loading,~\cite{Nott1994} we estimate that $\gamma_c \sim \left\langle \varepsilon \right\rangle^{2}$. According to Torquato et al.,~\cite{Torquato1990} $\left\langle \varepsilon \right\rangle \approx {(1-\phi)^3}\left\langle 2a \right\rangle/{12\phi(2-\phi)}$, where $\left\langle 2a \right\rangle$ is the average diameter of particles. Consequently, motion of suspending particles at the transition state can be evaluated by the evolution of $\gamma_c$ with $\phi$ as shown in Fig.\ref{Fig11} (a). $\gamma_c$ is deduced from the evolution of the viscosity with $\gamma$ in Fig.\ref{Fig7} for $30\;\%$ suspensions as well as Fig.\ref{Fig8} and Fig.\ref{Fig9} for $40\;\%$ and $50\;\%$ suspensions, respectively.

To determine the critical strain $\gamma_c$, we calculate the average value of the viscosity at the equilibrium state (with large strains) and the critical strain is determined at the location where the viscosity curves diverge from the horizontal solid lines standing for the equilibrium viscosity.~\cite{Lin2021} From the inset of Fig.\ref{Fig11} (a), it can be inferred that for G2, the particle motion approaches the convective limit, $\gamma_c \sim \left\langle \varepsilon \right\rangle$, while the behavior diverges from the convective motion, approaching $\gamma_c \sim \left\langle \varepsilon \right\rangle^2$ for G1. This indicates a more diffusive motion of the particles in the non-degassed suspension with glass particles, which reflects that the fine water droplets in the system disturb the convective motion for suspensions with smooth glass particles. As to the suspensions with PMMA particles, it is observed that $\gamma_c$ are similar for P1 and P2, both of which are larger than the one for the degassed suspension with smooth glass particles, G2. This may be due to roughness effect that disturbs the close contact of particles. Motion of particles in P1 are still close to $\gamma_c \sim \left\langle \varepsilon \right\rangle$ reflecting that the motion is nearly convective, which is not affected by the presence of the fine droplets. 


From the above investigation, we infer that water droplets behave differently in suspensions with glass particles and PMMA particles, thereby affecting the shear thinning of suspensions in different manners. The two types of suspending particles have a similar surface energy. Also, it is observed that the influence by particle roughness dominates over other properties of particles,~\cite{Lin2021a} therefore, we take particle roughness as the leading cause altering the interaction between particles and water droplets. The medium has a good wettability on the smooth glass particle due to its low surface tension with respect to the surface energy of the glass (about $40\;\text{mN/m}$),~\cite{Szymczyk2008} therefore, similar to the condition in microfluidic systems, water droplets are considered to move freely in the suspension,\cite{Guerrero2020,Giudice2021}, as shown in Fig.\ref{Fig11} (b). As a result, the droplet-particle interaction is considered to be repulsive, which prevents the close-contact interaction between solid particles. While, for suspensions with PMMA particles with considerable rough surfaces, we propose that droplets stick on the uneven surface of particles due to the contact line pinning effect, similar to the behavior of nano-bubbles on a solid surface ~\cite{Sun2016,Liu2013} as shown in Fig.\ref{Fig11} (c). As a result, the trapped droplets on particles enhance the adhesive interaction between particles via ``water bridging''. The particle cluster in shear flow undergoes a hydrodynamic moment estimated as, $M_H \sim l^3 \eta_0 \dot{\gamma}$, where $l$ is the length scale of the cluster. And the scale of the cluster formed is considered to be determined by a critical moment $M_c$, above which particles escape from the clusters to reduce $l$ to keep $M_H = M_c$.~\cite{Lin2021a} This leads to the decrease of $l$ with increasing $\dot{\gamma}$ and thereby the shear thinning behavior. Obviously, the magnitude of $M_c$ is determined by the adhesive strength between particles in the cluster. Considering the effect from water droplets, in G1, receding of the inter-particle interaction leads to a lower $M_c$, and therefore suppression of the shear thinning behavior. On the other hand, for P1, the enhancement on particle adhesion increases $M_c$, promoting the shear thinning phenomenon.

\begin{figure*}[tbh]
 \includegraphics[width=0.8\textwidth]{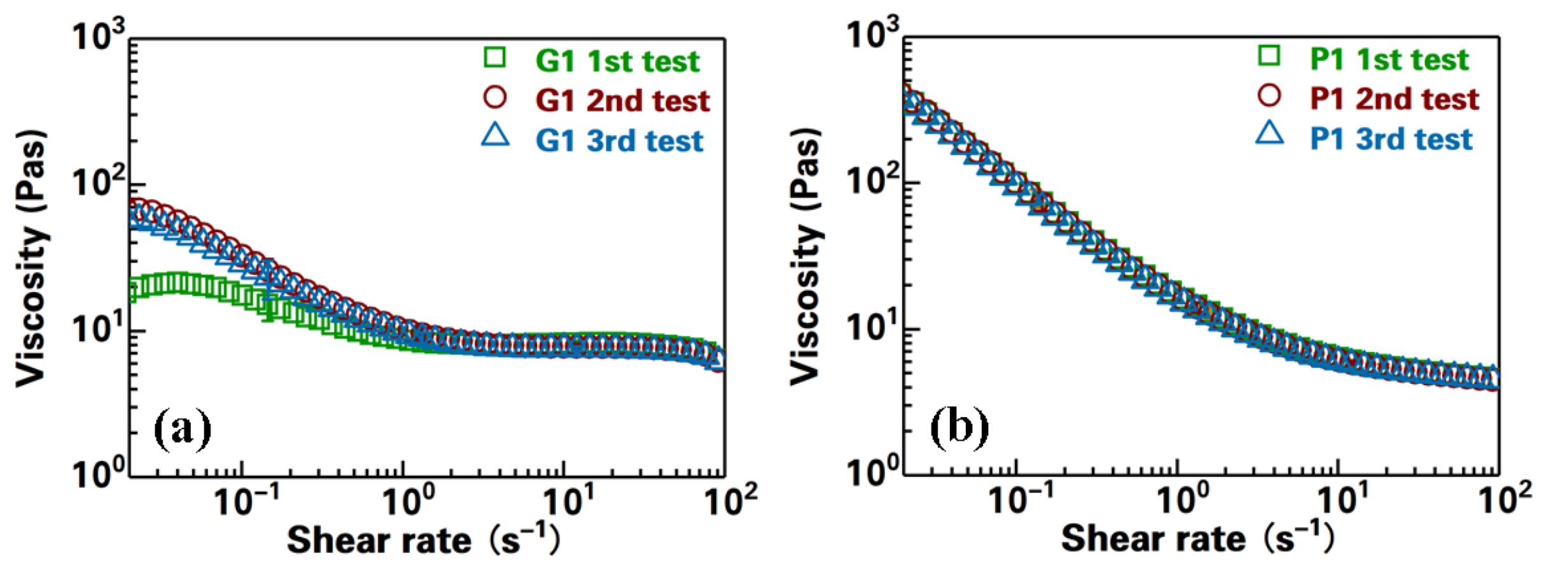}\\
  \caption{Repeatability test on a single sample of the non-degassed suspension with $40\;\%$ (a) glass particles and (b) PMMA particles, respectively, at an artificial relative humidity above $75 \%$ at $25\;^\circ\text{C}$}\label{Fig12}
\end{figure*}

Similar to the droplets in micro- or nano-emulsions, the fine droplets are observed to be stable at the static state,~\cite{Ankur2016,McClements2012} the influence of which on the rheology of suspensions is not receded after resting for $3$ days. Nevertheless, it is found that droplets in G1 are sensitive to shear. The shear thinning curve of G1 approaches that of G2 when the ramp-up shear test is applied for more than once, as shown in Fig.\ref{Fig11} (a). While for P1, we find that the shear thinning is still reproducible after several tests, as shown in Fig.\ref{Fig12} (b). Therefore, we discern that droplets are more stable attaching on the particles than at the free state. This also confirms the distinct droplet-particle interactions in G1 and P1.

\section{CONCLUSION}

In this study, we have shown that the shear thinning behavior of the oil-based non-Brownian suspension is sensitive to the ambient condition because of the moisture present during the production of non-Brownian suspensions in the laboratory. We propose that this occurs due to the introduction of small water droplets into the oil-based suspension, which influences the motion of suspending particles, and causes increasing or decreasing interaction forces between particles depending on their surface roughness. Consequently the shear-induced particle aggregation is either promoted or suppressed, leading to the variation of the shear thinning of the suspension. Our results put into question many, if not all, laboratory studies on the flow rheology of non-Brownian suspensions, since they usually do not report to have dehumidified the ambient air prior the suspension production and/or degassed the suspension afterward.~\cite{Singh2003,Dai2013,Lin2014,Denn2014,Tanner2018}


\begin{acknowledgments}
This study is supported by National Natural Science Foundations of China (No. 41976055 and 12272344), and Zhejiang Provincial Natural Science Foundation of China (Grant No. LY20A020008), Finance Science and Technology Project of Hainan Province (Grant No. ZDKJ202019) and 2020 Research Program of Sanya Yazhou Bay Science and Technology City (Grant No. SKYC-2020-01-001).
\end{acknowledgments}

\section*{DATA AVAILABILITY}
The data that support the findings of this study are available from the corresponding authors upon reasonable request.


%

\end{document}